\documentclass[conference]{IEEEtran}
\usepackage[latin9]{inputenc}
\usepackage{geometry}
\usepackage{color}
\usepackage{array}
\usepackage{float}
\usepackage{booktabs}
\usepackage{multirow}
\usepackage{amsmath}
\usepackage{amssymb}
\usepackage{graphicx}

\makeatletter

\providecommand{\tabularnewline}{\\}

\IEEEoverridecommandlockouts
\usepackage{cite}
\usepackage{amsfonts}\usepackage{algorithmic}
\usepackage{textcomp}
\usepackage{xcolor}
\def\BibTeX{{\rm B\kern-.05em{\sc i\kern-.025em b}\kern-.08em
    T\kern-.1667em\lower.7ex\hbox{E}\kern-.125emX}}

\usepackage{subcaption}
\usepackage{enumerate}

\makeatother

\begin{document}
\title{A Data Quality Assessment Framework for AI-enabled Wireless Communication
\thanks{Corresponding author: Rui Zhou and Zhi-Quan Luo. This work was supported
by Huawei Research Grant.} }
\author{\IEEEauthorblockN{Hanning Tang$^{\dag}$, Liusha Yang$^{\dag}$, Rui Zhou$^{\dag}$,
Jing Liang$^{\ddag}$, Hong Wei$^{\ddag}$, Xuan Wang$^{\ddag}$,
\\Qingjiang Shi$^{\S\dag}$, and Zhi-Quan Luo$^{\dag}$} \\ \IEEEauthorblockA{$^{\dag}$Shenzhen Research Institute of Big Data, The Chinese University
of Hong Kong, Shenzhen, China\\$^{\ddag}$Wireless Research Department,
Huawei Company, Shanghai, China\\$^{\S}$School of Software Engineering,
Tongji University, Shanghai, China}}
\maketitle
\begin{abstract}
Using artificial intelligent (AI) to re-design and enhance the current
wireless communication system is a promising pathway for the future
sixth-generation (6G) wireless network. The performance of AI-enabled
wireless communication depends heavily on the quality of wireless
air-interface data. Although there are various approaches to data
quality assessment (DQA) for different applications, none has been
designed for wireless air-interface data. In this paper, we propose
a DQA framework to measure the quality of wireless air-interface data
from three aspects: \emph{similarity}, \emph{diversity}, and \emph{completeness}.
The similarity measures how close the considered datasets are in terms
of their statistical distributions; the diversity measures how well-rounded
a dataset is, while the completeness measures to what degree the considered
dataset satisfies the required performance metrics in an application
scenario. The proposed framework can be applied to various types of
wireless air-interface data, such as channel state information (CSI),
signal-to-interference-plus-noise ratio (SINR), reference signal received
power (RSRP), etc. For simplicity, the validity of our proposed DQA
framework is corroborated by applying it to CSI data and using similarity
and diversity metrics to improve CSI compression and recovery in Massive
MIMO systems.
\end{abstract}


\section{Introduction \label{sec: Introduction}}

Nowadays, data has become ubiquitous with the development of modern
information technologies. Various applications based on the extraction
of meaningful information from data have been studied. However, the
data quality is not self-evident due to reasons such as unreliable
sources or errors injected when data is transferred or stored \cite{saha2014data}.
When applications are fed with the low-quality data, the obtained
decisions may become unreliable and mistaken. Therefore, the data
quality assessment (DQA) must be conducted to evaluate and help to
improve the data quality \cite{sidi2012data}. Generally, the goal
of DQA is to check whether the dataset on hand is fit to be used for
a specified task. The detailed assessment process depends on the properties
of data and specific applications.

The DQA is conducted by measuring the properties of given data in
terms of several interested criteria, which may vary with the data
types and their corresponding tasks.  A comprehensive survey of data
quality criteria is presented in \cite{sidi2012data}.  We introduce
several traditionally adopted data criteria in the following \cite{cichy2019overview}: 
\begin{itemize}
\item \emph{Accuracy}: the extent to which data are correct, reliable and
certified. 
\item \emph{Timeliness}: the extent to which the age of the data is appropriate
for the task at hand. 
\item \emph{Consistency}: the extent to which data are presented in the
same format and compatible with previous data. 
\item \emph{Accessibility}: the extent to which information is available,
or easily and quickly retrievable. 
\end{itemize}
A comprehensive DQA result is usually obtained by combining all considered
criteria measuring results \cite{pipino2002data}.

In this paper, we are particularly interested in studying DQA in the
context of AI-enabled wireless communications \cite{8808168,9237460}.
More specifically, we propose to build a DQA framework for the wireless
air-interface data, whose quality is essential for the performance
of the AI algorithms used in wireless communication networks. Note
that those problems focused on by the traditional DQA are assumed
to be handled in the pre-processing stage, which usually performs
data cleaning process to ensure that data is correct, consistent and
usable. In this paper, we propose a specific DQA framework for AI-enabled
wireless communications with tailored data criteria in order to facilitate
the AI algorithms to make full use of data and improve their ultimate
performance.

To this end, the major goal of this paper is to develop a DQA framework
for AI-enabled wireless communications. It consists of three quality
criteria\footnote{The detailed discussion on completeness is omitted due to the page
limit. It will be included in the future journal version of this work.}:
\begin{itemize}
\item \emph{Similarity}: the extent to which two datasets are close to each
other. A high similarity measuring result indicates that the difference
between two considered datasets is small.
\item \emph{Diversity}: the extent to which data are rich and diverse. A
high diversity measuring result indicates that the value of embedded
information is large.
\item \emph{Completeness}: the extent to which the considered data satisfy
the required performance metrics in an application scenario.
\end{itemize}
The similarity criterion is useful in merging and clustering datasets.
For example, we can merge datasets admitting large similarity to augment
a small-sized dataset so that it can be used with applications requiring
a large number of samples. The diversity criterion is useful in estimating
the generalization ability of trained AI models. Intuitively, if a
model is trained by more diverse data, the obtained model is likely
to be of good performance in a broader range of scenarios and even
unseen ones.

In this paper, we first give the introduction in Sec. \ref{sec: Introduction}.
Then we present the \textcolor{black}{detailed similarity and diversity
measurement processes in Sec. \ref{sec: similarity} and Sec. \ref{sec: diversity}.
In Sec. \ref{sec: similarity on CSI data} and Sec. \ref{sec: diversity on CSI data},
we validate the the proposed DQA framework on CSI data. Finally, the
conclusion is given in Sec. \ref{sec: Conclusion}.}

\section{Similarity \label{sec: similarity}}

The overall process of measuring the similarity between two datasets
can be summarized into the following four steps. Note that the selection
of methods in each step should depend on the specific data type and
application.
\begin{enumerate}
\item \textbf{\emph{Feature extraction (optional)}}: extract meaningful
feature samples from the original datasets; 
\item \textbf{\emph{Inter-set distance}}: compute the distance between each
pair of samples belonging to two different datasets; 
\item \textbf{\emph{Dataset difference}}: compute the difference between
two sample sets using the obtained distances; 
\item \textbf{\emph{Aggregation}}: summarize all similarity measuring results. 
\end{enumerate}

\subsection{Feature Extraction \label{subsec: Feature-extraction}}

Feature extraction starts with a set of sampled data and produces
derived values (features) that are informative and non-redundant.
Measuring the similarity of features extracted from original datasets
may yield more interpretable results. There are various methods for
extracting features, such as Fourier transformation, wavelets transformation,
filter, convolutional neural network, principal component analysis,
etc.

\subsection{Inter-set Distance \label{subsec: Samples distance}}

There are many options for measuring the distance between two samples.
Denote by $\mathbf{x}$ and $\mathbf{y}$ ($\mathbf{x},\mathbf{y}\in\mathbb{C}^{N}$)
the two samples. We consider the following distance measures \cite{deborah2015comprehensive}: 
\begin{itemize}
\item Euclidean distance: 
\begin{equation}
d_{Eu}(\mathbf{x},\mathbf{y})=\|\mathbf{x}-\mathbf{y}\|_{2};\label{eq: eu definition}
\end{equation}
\item Geman McClure (GMC) distance: 
\begin{equation}
d_{GMC}(\mathbf{x},\mathbf{y})=\sum_{i=1}^{N}\frac{|x_{i}-y_{i}|^{2}}{1+|x_{i}-y_{i}|^{2}};
\end{equation}
\item Euclidean distance of cumulative spectrum (ECS) distance (only for
$\mathbf{x},\mathbf{y}\in\mathbb{R}^{N}$): 
\begin{equation}
d_{ECS}(\mathbf{x},\mathbf{y})=\|\mathbf{c}_{\mathbf{x}}-\mathbf{c}_{\mathbf{y}}\|_{2},\label{eq: ECS definition}
\end{equation}
where $\mathbf{c}_{\mathbf{x}}$, $\mathbf{c}_{\mathbf{y}}\in\mathbb{R}^{N}$
are \textcolor{black}{the cumulative summation of $\mathbf{x}$ and
$\mathbf{y}$,} i.e., $c_{\mathbf{x},i}=\sum_{k=1}^{i}x_{k}$, $c_{\mathbf{y},i}=\sum_{k=1}^{i}y_{k}$.
Especially, when $\mathbf{X}$,$\mathbf{Y}\in\mathbb{R}^{N_{1}\times N_{2}}$
are two matrices, then the ECS distance between $\mathbf{X}$ and
$\mathbf{Y}$ is similarly defined as $d_{ECS}(\mathbf{X},\mathbf{Y})=\|\mathbf{C}_{\mathbf{X}}-\mathbf{C}_{\mathbf{Y}}\|_{F},$
where $\mathbf{C}_{\mathbf{X}}$, $\mathbf{C}_{\mathbf{Y}}\in\mathbb{R}^{N_{1}\times N_{2}}$
with $C_{\mathbf{X},ij}=\sum_{l=1}^{i}\sum_{k=1}^{j}X_{lk}$ and $C_{\mathbf{Y},ij}=\sum_{l=1}^{i}\sum_{k=1}^{j}Y_{lk}$. 
\end{itemize}
There are many measures that are not introduced here, e.g., Jeffrey
divergence, cosine similarity, Pearson $\mathcal{X}^{2}$ distance,
and squared chord distance, due to limited space in this paper.

\subsection{Dataset Difference \label{subsec: Datasets difference}}

When considering the similarity between two datasets, the underlying
distributions of these datasets are essential for determining their
similarity. The similarity can be measured via the difference between
the underlying distributions, i.e., the smaller the difference, the
higher the similarity.

Given two random variables $X,Y\in\mathbb{C}^{N}$, we should technically
measure their difference using their probability distribution. But
in practice we can only get access to $\mathcal{X}=\{\mathbf{x}_{i}\}_{i=1}^{n_{x}}$
and $\mathcal{Y}=\{\mathbf{y}_{i}\}_{i=1}^{n_{y}}$, which are two
datasets of samples sampled from them. Their underlying distributions
are unknown to us. Therefore, we can only estimate their difference
by their empirical distribution.

Assume $\mathcal{X}=\{\mathbf{x}_{i}\}_{i=1}^{n_{x}}$, $\mathcal{Y}=\{\mathbf{y}_{i}\}_{i=1}^{n_{y}}$
are i.i.d.\footnote{Different sample strategies are also allowed. But one needs to use
the corresponding estimation method.} samples from $X$, $Y$, respectively. We consider the following
distance measures: 
\begin{itemize}
\item \emph{Mean distance}: a simple approach is estimating the mean distance
between $\mathbf{x}\sim X$ and $\mathbf{y}\sim Y$, i.e.,
\begin{equation}
\mathbb{E}_{\mathbf{x}\sim X,\mathbf{y}\sim Y}[d(\mathbf{x},\mathbf{y})]\approx\frac{1}{n_{x}n_{y}}\sum_{i=1}^{n_{x}}\sum_{j=1}^{n_{y}}d(\mathbf{x}_{i},\mathbf{y}_{j});\label{eq: mean definition}
\end{equation}
where $d(\cdot,\cdot)$ is a distance measure mentioned in Sec. \ref{subsec: Samples distance}. 
\item \emph{Maximum mean discrepancy}: a biased empirical estimate of maximum
mean discrepancy (MMD) is 
\begin{align}
MMD\left(\mathcal{X},\mathcal{Y}\right)=[ & \frac{1}{n_{x}^{2}}\sum_{i,j=1}^{n_{x}}k(\mathbf{x}_{i},\mathbf{x}_{j})\\
 & -\frac{2}{n_{x}n_{y}}\sum_{i,j=1}^{n_{x},n_{y}}k(\mathbf{x}_{i},\mathbf{y}_{j})\\
 & +\frac{1}{n_{y}^{2}}\sum_{i,j=1}^{n_{y}}k(\mathbf{y}_{i},\mathbf{y}_{j})]^{1/2},\label{eq: mmd definition}
\end{align}
where $k(\mathbf{x},\mathbf{y})=\exp(-\frac{d(\mathbf{x},\mathbf{y})^{2}}{2\sigma^{2}})$
is often selected and $d(\cdot,\cdot)$ is a distance measure. MMD
is widely used in domain adaptation \cite{long2016unsupervised,rozantsev2018beyond}
and generative adversarial networks \cite{binkowski2018demystifying,li2017mmd}. 
\item \emph{Leave-one-out accuracy of nearest neighbor classifier (NNCA)}:
the 1-Nearest Neighbor (1-NN) classifier is used in two-sample tests
to assess whether two distributions are identical \cite{lopez2016revisiting}.
Assume that samples in $\mathcal{X}$ are labeled with positive and
samples in $\mathcal{Y}$ are labeled with negative, then the accuracy
of this classifier is defined as 
\begin{equation}
\text{Accuracy}=\frac{TP+TN}{n_{x}+n_{y}},
\end{equation}
where $TP$ is the true positive number and $TN$ is the true negative
number of the leave-one-out test results from the 1-NN classifier.
The distance function used in 1-NN classifier is one of methods mentioned
in Sec. \ref{subsec: Samples distance} The 1-NN classifier should
yield a near $50\%$ accuracy when the two datasets are very similar,
while a near $100\%$ accuracy when the two datasets are very different. 
\item \emph{Wasserstein distance}: the Wasserstein distance ($W_{p}$) \cite{villani2003topics},
a.k.a. optimal transport distance, is computed as 
\begin{equation}
W_{p}(\mathcal{X},\mathcal{Y})=(\min_{\mathbf{T}\in{\cal U}}\text{Tr}(\mathbf{D}_{p}^{T}\mathbf{T}))^{1/p},\label{eq: wp definition}
\end{equation}
where $i,j$-th element of $\mathbf{D}_{p}$ is $d(\mathbf{x}_{i},\mathbf{y}_{j})^{p}$,
$d(\cdot,\cdot)$ is a distance mentioned in Sec. \ref{subsec: Samples distance},
$p\ge1$, and ${\cal U}=\{\mathbf{T}\in\mathbb{R}^{n_{x}\times n_{y}}\vert\sum_{i=1}^{n_{x}}\mathbf{T}_{ij}=\frac{1}{n_{y}},\sum_{j=1}^{n_{y}}\mathbf{T}_{ij}=\frac{1}{n_{x}},\mathbf{T}_{ij}\ge0\}$.
Calculating this distance corresponds to solving a linear programming
problem, which can be efficiently done by off-the-shelf solvers \cite{xie2020fast}. 
\end{itemize}
There are also measures that are not introduced here due to limited
space, such as $f$-divergence, total variation distance, integral
probability metrics, etc. It should be noted that each of these measures
has its own distinct properties and should be chosen based on the
specific applications. 

\subsection{Aggregation \label{subsec: Aggregation}}

Summary methods such as \emph{minimum}, \emph{maximum}, or \emph{weighted
average} operations can be used to handle the aggregation of the similarity
of multiple features extracted from datasets \cite{pipino2002data}.
One can compute the minimum (or maximum) value of the normalized similarity
of the individual features. The minimum operator is conservative in
that it assigns an aggregate value no higher than the value of its
weakest similarity (normalized to between 0 and 1). If one has a good
understanding of the importance of each features to the overall evaluation
of similarity, for example, then a weighted average is appropriate.
To ensure the similarity is normalized, each weighting factor should
be between zero and one, and the weighting factors should add to one.

\section{Diversity \label{sec: diversity}}

Diversity is defined as the richness and evenness of the considered
dataset. The data diversity measurement consists of three steps: \emph{1)
feature extraction; 2) Intra-set Distance; 3) dataset diversity measurement;
4) aggregation}. The first and the fourth steps follow the same procedures
as introduced in Sec. \ref{subsec: Feature-extraction} and Sec. \ref{subsec: Aggregation}.
The second step is also similar to Sec. \ref{subsec: Samples distance}
but computes the distance between each pair of samples belonging to
same dataset.  We present potential methods for the third step as
follows.
\begin{enumerate}
\item \textit{Entropy-based method}\textit{\emph{: }}if features are scalars,
we propose to directly (skip the second step) use Shannon entropy
to compute their diversity. Given a dataset ${\mathcal{X}}=\{{x}_{i}\}_{i=1}^{n}$
where $x_{i}\in\mathbb{R}$ is a scalar, we first obtain the empirical
distribution of ${\mathcal{X}}$ where the support is divided into
$S$ bins, and the diversity is further computed as 
\begin{equation}
D_{s}(\mathcal{X})=-\sum_{i=1}^{S}p_{i}\log{p_{i}}/\log{S},\label{div_shannon}
\end{equation}
where $p_{i}$ is the empirical probability of samples in the $i$-th
bin. Here $S$ can be adjusted according to the practice and the bin
width can be either uniform or manually designed. 
\item \textit{Distance-based method \cite{gong2019diversity}}\textit{\emph{:}}
given a feature dataset ${\mathcal{Y}}=\{{\bf y}_{i}\}_{i=1}^{n}$,
where ${\bf y}_{i}\in\mathbb{C}^{N}$ is a vector (or a matrix after
vectorization), the diversity of ${\mathcal{Y}}$ is computed as 
\begin{equation}
D_{a}(\mathcal{Y})=\frac{1}{n(n-1)/2}\sum_{i\neq j}^{n}d(\mathbf{y}_{i},\mathbf{y}_{j}).\label{eq: avg_div}
\end{equation}
where distance $d(\cdot,\cdot)$ can be one of the sample distance
measures mentioned in Sec. \ref{subsec: Samples distance}. 
\item \textit{Determinantal point process (DPP)-based method}\emph{: }inspired
by the definition of DPP \cite{macchi1975coincidence}, the diversity
of ${\mathcal{Y}}$ can be computed by $D_{d}(\mathcal{Y})=\mathrm{det}(\mathbf{L})$
where $\mathrm{det}(\cdot)$ denotes the determinant of matrix and
$\mathbf{L}$ is a positive semidefinite kernel matrix where $L_{ij}$
is the pairwise kernel function value of ${\bf y}_{i}$ and ${\bf y}_{j}$.
For example, $L_{ij}$ can be the radial basis function kernel, i.e.,
$L_{ij}=-\exp\left(-\frac{\lVert\mathbf{y}_{i}-\mathbf{y}_{j}\rVert^{2}}{2\sigma^{2}}\right)$
where $\sigma$ is a hyperparameter. 
\item \textit{Compression-based method}: inspired by \cite{deng2009imagenet}
and \cite{mithun2019construction} that evaluate the diversity of
image datasets, we propose to use the method based on image compression
to measure the diversity of $\mathcal{Y}$. It first simply computes
the sample mean of data in $\mathcal{Y}$, i.e., $\bar{{\bf y}}=\frac{1}{n}\sum_{i=1}^{n}{\bf y}_{i}$,
and then turns $\bar{{\bf y}}$ into a grayscale image and saves it
as a JPG file. The inverse of the size of JPG file represents the
diversity of considered dataset. The idea is that a diverse dataset
will result in a blurrier average image, which has less information
and therefore a smaller JPG file size.
\end{enumerate}
It is claimed that the proposed general DQA framework can be used
with all types of wireless air-interface data. Therefore, to illustrate
the usage of our proposed framework, we give an example of applying
the proposed DQA framework to the CSI data in the next.

\section{Applying Similarity Measure to CSI Data \label{sec: similarity on CSI data}}

\subsection{Selection of Methods \label{subsec: method selection of similarity}}

To measure the similarity of CSI dataset, we apply the proposed DQA
framework described in Sec. \ref{sec: similarity}. In the feature
extraction step, we use classical Fourier transformation to extract
the the power delay profile (PDP), Doppler, and the angular power
spectrum (APS) \cite{tse2005fundamentals,yin2020addressing}. The
sparsity of PDP, Doppler and APS can be further extracted using the
Hoyer method \cite{hurley2009comparing}. Next, we compute the samples
distance of each pair of features using ECS distance as recommended
in \cite{deborah2015comprehensive}. Then, we consider the datasets
difference measures mentioned in Sec. \ref{subsec: Datasets difference}.
Finally, we aggregate the similarity of different features by average. 

Here we illustrate the comparison of the proposed similarity measures
for CSI datasets via experiment results. We generate a group of synthetic
CSI datasets using QuaDRiGa \cite{jaeckel2014quadriga} by keeping
the RMS delay spread range the same ($2000\,ns$) for each dataset
but change their offset.  The details of generating these datasets
are described in Appendix \ref{APX: similarity datasets}. Fig. \ref{fig: datasets similarity}
shows the normalized differences by applying Mean distance, MMD, NNCA
and $W_{p}$ ($p=2$) methods on PDP feature. $\Delta_{rmsDSoffset}$
is the difference between the offset of the RMS delay spread range
settings of each pair of datasets.  It is significant that only the
results of $W_{p}$ have the desired linear response \cite{deborah2015comprehensive}.
The results of NNCA also have the linear response before saturation,
i.e., when $\Delta_{rmsDSoffset}\le2000\,ns$. It is consistent with
the range of the RMS delay spread setting. Therefore, the details
of selected methods are described in Table \ref{tab: methods for similarity - CSI}.

\begin{table}[H]
\centering{}\caption{Selection of methods in similarity of CSI data. \label{tab: methods for similarity - CSI}}
\begin{tabular}{lcc}
\hline 
~~~~~~~~~~~Step & Method (Sparsity) & Method (Others)\tabularnewline
\hline 
Feature extraction & Hoyer & FT\tabularnewline
Inter-set distance & \multicolumn{2}{c}{ECS}\tabularnewline
Dataset difference & \multicolumn{2}{c}{$W_{p}$/NNCA}\tabularnewline
Aggregation & \multicolumn{2}{c}{Average}\tabularnewline
\hline 
\end{tabular}
\end{table}

\begin{figure}
\centering \includegraphics[width=1.05\linewidth]{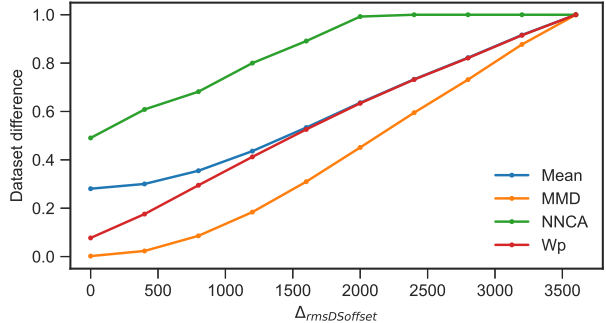}
\caption{Normalized differences from Mean distance, MMD, NNCA and $W_{p}$
($p=2$) methods in terms of PDP features versus the difference of
the RMS delay spread setting. \label{fig: datasets similarity}}
\end{figure}

\subsection{Data Augmentation \label{sec:data_aug}}

In this subsection, we consider to augment the small-sized training
dataset for CsiNet algorithm \cite{wen2018deep} by using the similarity
measure. Well training a neural network, e.g., CsiNet, for a certain
scenario requires a large amount of samples from that specific scenario.
But sampling all the training data on-site is too expensive to be
practical. Therefore, we propose to augment a small-sized dataset
by merging a few candidate datasets (perhaps generated using synthetic
data platform, e.g., QuaDRiGa) with the reference of our proposed
similarity measures. The detailed steps of the proposed augmentation
process are as follows: 
\begin{enumerate}
\item obtain a (probably small-sized) dataset $\tilde{\mathcal{Y}}$ from
the particular scenario; 
\item calculate the similarity between $\tilde{\mathcal{Y}}$ and all candidate
datasets; 
\item select $k$ candidate datasets most similar to $\tilde{{\cal {Y}}}$
and combine them together as a training dataset, where $k$ can be
determined by the budget or a threshold of the similarity. 
\end{enumerate}
To illustrate the performance of our proposed method, we generate
a candidate dataset pool containing 100 datasets $\mathcal{X}_{1},\mathcal{X}_{2},\dots,\mathcal{X}_{100}$.
Each of them consists of 100 samples generated by the CDL model. A
test dataset $\mathcal{Y}$ is generated by the Urban Macro-Cell (UMa)
model. The details of generating the above datasets are described
in Appendix \ref{APX: test data for similarity}. The reference dataset
$\tilde{\mathcal{Y}}$ contains only 100 samples randomly selected
from test dataset $\mathcal{Y}$.

During the training process, the mean squared error loss (MSE) function
and the default adaptive momentum optimizer are adopted with epochs,
learning rate, batch size and data compression ratio set as $100$,
$0.001$, $128$, and $1/4$. The input of CsiNet requires ${\bf H}$
to be transformed to delay domain through discrete Fourier transform,
which is denoted as $\tilde{{\bf H}}$. The performance of CsiNet
is evaluated by a normalized MSE (NMSE) between the recovered $\hat{{\bf H}}$
and original $\tilde{{\bf H}}$, defined as $\text{NMSE}=\mathbb{E}\left[\frac{\|\tilde{{\bf H}}-\hat{\mathbf{H}}\|_{2}^{2}}{\|\tilde{{\bf H}}\|_{2}}\right]$.
In the test phase, we use $W_{p}$ ($p=2$) and NNCA to measure the
difference between datasets, and the differences between PDP, APS,
PDP sparsity, APS sparsity and the average difference of these four
features are used to construct the training dataset.

\begin{figure}
\raggedright{}\includegraphics[scale=0.62]{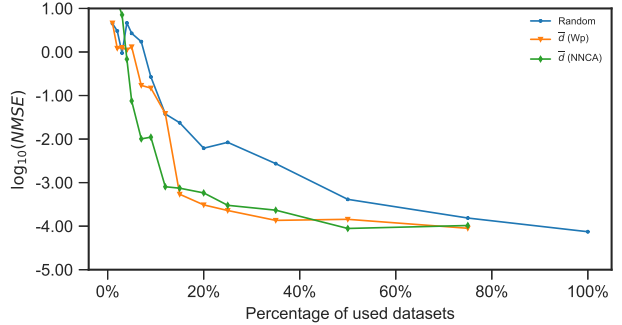}\caption{Performance of trainded CsiNet. $\overline{d}$ is the average difference
of PDP, APS and thier sparsity. \label{fig: data augmentation}}
\end{figure}

As shown in Fig. \ref{fig: data augmentation}, when the CsiNet algorithm
is fed with top 25\% of candidate datasets most similar to $\tilde{\mathcal{Y}}$,
the performance of the trained CsiNet is already close to that of
the network trained with the whole candidate dataset pool. As a comparison,
the NMSE of CsiNet trained by the randomly selected datasets shows
significantly worse performance when only 25\% of the candidate datasets
are used.  It means that our proposed method can augment a small
sampled dataset in an efficient and reasonable way, so that the performance
of CsiNet can be quite good with only a fraction of the whole dataset.
Thus in practical application, the cost of sampling a real dataset
and the following model training are expected to be dramatically reduced.

\section{Applying Diversity Measure to CSI Data \label{sec: diversity on CSI data}}

\subsection{Selection of Methods}

To measure the diversity of the CSI dataset, we apply the proposed
DQA framework described in Sec. \ref{sec: diversity}. Firstly, we
obtain data features in the same way as described in Sec. \ref{subsec: method selection of similarity}.
Then we compute the diversities of each features. Different diversity
measures may be adopted for different features. For the sparsity features,
we choose the entropy-based method. For the PDP, Doppler and APS,
the distance-based, the DPP-based and the compression-based methods
may be used. Finally, we yield the diversity evaluation result of
the CSI dataset by averaging the feature diversities.

\begin{figure}
\centering \includegraphics[width=0.87\linewidth]{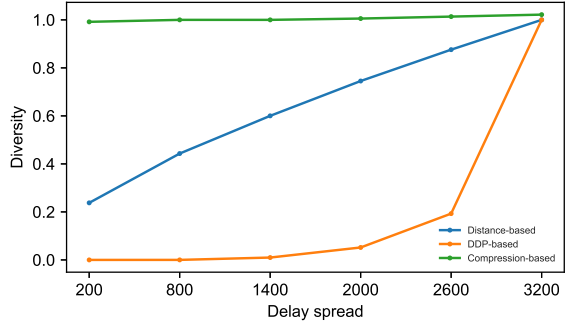}
\caption{Normalized diversities computed by distance-based, DDP-based and compression-based
methods.}
\label{fig: PDPdiv}
\end{figure}

In the following, we conduct an experiment to compare the performance
of distance-based, the DPP-based and the compression-based methods
on measuring the PDP diversity of CSI datasets. We generate $6$ CSI
datasets with RMS delay spread ranging from $20\,ns$ to $3200\,ns$.
The other settings are the same as described in Appendix \ref{APX: similarity datasets}.
 In the distance-based method, similar to that in Sec. \ref{subsec: method selection of similarity},
we use ECS to measure the distances between features. As in Fig. \ref{fig: PDPdiv},
the diversity obtained by the distance-based method is almost a linear
function of the RMS delay spread, while the diversity computed by
the DDP-based method increases sharply when the delay spread gets
large. Since the dimension of PDP feature is not sufficiently large,
the sizes of the JPG files  after compression are all quite small
and their differences are not significant.  We obtain similar experiment
results for Doppler and APS features, which is not present due to
the page limit.  Therefore, the details of selected methods are described
in Table \ref{tab: methods for diversity - CSI}.

\begin{table}[H]
\centering{}\caption{Selection of methods in diversity of CSI data. \label{tab: methods for diversity - CSI}}
\begin{tabular}{lcc}
\hline 
~~~~~~~~~~~Step & Method (Sparsity) & Method (Others)\tabularnewline
\hline 
Feature extraction & Hoyer & FT\tabularnewline
Intra-set distance & \multirow{2}{*}{Entropy-based method} & ECS\tabularnewline
Dataset diversity &  & Distance-based method\tabularnewline
Aggregation & \multicolumn{2}{c}{Average}\tabularnewline
\hline 
\end{tabular}
\end{table}

\subsection{Predicting the Generalization Power of Models \label{sec:data_div} }

The diversity of training data is essential to the machine learning
applications. Intuitively, if a model is trained by more diverse data,
the obtained model is likely to be of good performance in a broader
range of scenarios and even unseen ones. Here we consider the application
of data diversity in the training of the CsiNet model. 

We generate $84$ training datasets through the CDL model by QuaDRiGa,
with each of them containing $5000$ samples.  Detailed descriptions
of the generation settings of these datasets are given in Appendix
\ref{APX: test data for diversity}. Since the CSI data fed to CsiNet
contains only one time interval, Doppler and its sparsity features
can not be extracted. We compute the overall diversity $\tilde{D}$
of each training dataset by averaging the diversities of PDP, APS,
PDP sparsity and APS sparsity (denoted by $D_{{\rm PDP}}$, $D_{{\rm APS}}$,
$D_{{\rm PDPspar}}$ and $D_{{\rm APSspar}}$), i.e., $\tilde{D}=\frac{1}{4}(D_{{\rm PDP}}+D_{{\rm APS}}+D_{{\rm PDPspar}}+D_{{\rm APSspar}})$.
The test dataset consists of $2000$ samples, which should be more
diverse than the training datasets. It is generated through the UMa
model with $20\%$ outdoor UEs and $80\%$ indoor UEs.\footnote{Since the entries in $\tilde{{\bf H}}$ generated by the CDL model
are not of the same order of magnitude as those in $\tilde{{\bf H}}$
generated by the UMa model, we normalize $\tilde{{\bf H}}$ by dividing
with its maximum entry in both the training and the test datasets.} The same settings in Table \ref{tab: simulation parameters} are
used and other parameters follow their default values. Then $84$
trained CsiNets are obtained using $84$ different training datasets.

\begin{figure}
\centering{}\includegraphics[width=0.9\linewidth]{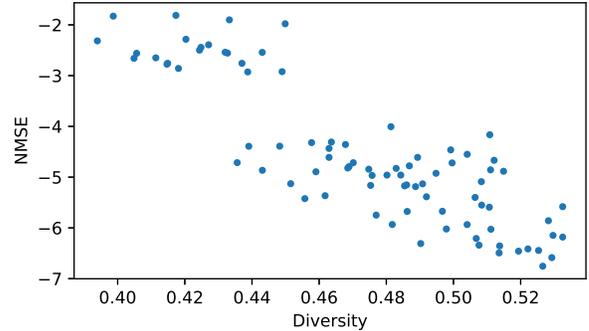}
\vspace{-0.3cm}
 \caption{Diversity of training datasets versus NMSE.}
\label{fig:div_NMSE}
\end{figure}

Fig. \ref{fig:div_NMSE} shows the diversity of the training dataset
versus NMSE on the test data. As we can see, NMSE decreases with the
increase of data diversity. This clearly demonstrates that, by increasing
the diversity of training data, the trained model could achieve superior
performance on a most diverse test dataset. If we would like to train
a network with good performance in a wide range of data scenarios,
the diversity of the training dataset can be a preliminary reference
before model training.

\section{Conclusion \label{sec: Conclusion}}

In this paper, we have proposed a general DQA framework for the AI-enabled
wireless communications, which to our knowledge, has not been developed
before. The currently proposed DQA framework consists of three quality
criteria: \emph{similarity}, \emph{diversity}, and \emph{completeness}.
We have presented a detailed framework structure for measuring the
similarity and diversity, and have shown the application of our proposed
DQA framework to the CSI data. The significant results of using our
proposed similarity and diversity measures in merging and evaluating
datasets have corroborated their validity. Future promising research
directions include generalizing this DQA framework for other types
of wireless air-interface data and exploring more meaningful quality
criteria.

\appendix
{}

\subsubsection{Datasets Used in Section \ref{sec: similarity on CSI data}\label{APX: similarity datasets}}

A group of CSI datasets generated by the CDL model are used. The basic
parameters are shown in Table \ref{tab: simulation parameters}, and
the antenna model of BS is 3GPP-MMW. Each group has 10 datasets, and
each dataset has 200 samples. Each dataset includes CSI samples with
RMS delay spread uniform sampled from {[}$y$, $y+2000${]} (the user
speed and path angles are fixed), where $y$ of each dataset varies
from 0 to 3600 $ns$.

\begin{table}[htbp]
\centering{}\caption{Basic simulation parameters. \label{tab: simulation parameters}}
\begin{tabular}{cc}
\toprule 
Parameter & Value\tabularnewline
\midrule 
$f_{c}$ & 2.16GHz\tabularnewline
$B$ (Band width) & 20MHz\tabularnewline
$f_{0}$ & 60KHz\tabularnewline
$N_{f}$ & 52\tabularnewline
$N_{T},(N_{v},N_{h})$ & 64, (8,8)\tabularnewline
$N_{R}$ & 1\tabularnewline
$f_{s}$ (Sample frequency) & 200Hz\tabularnewline
\bottomrule
\end{tabular}
\end{table}

\subsubsection{Datasets Used in Section\,\ref{sec:data_aug}\label{APX: test data for similarity}}

The common settings of the candidate training datasets pool $\mathcal{X}_{1},\mathcal{X}_{2},\dots,\mathcal{X}_{100}$
and test datasets $\mathcal{Y}$ are the same with that described
in Appendix \ref{APX: similarity datasets}, except that the BS is
set to have $16$ antennas, with $2$ rows and $8$ columns. For the
generation of $\mathcal{X}_{1},\mathcal{X}_{2},\dots,\mathcal{X}_{100}$,
the parameter range of path number, time delay, AOD, and ZOD of each
dataset are randomly generated as follows: 
\begin{itemize}
\item path number: $[\max(0,\lfloor n_{p}\rfloor-2),\lfloor n_{p}\rfloor+5]$,
where $n_{p}\sim U([1,10])$ ($U([a,b])$ is the uniform distribution
on {[}a,b{]}); 
\item time delay (ns): $[\max(0,n_{t}-\frac{w}{2}),n_{t}+\frac{w}{2}]$,
where $t\sim U([0,2500])$, $w$ are uniformly random selected from
\{100, 200, 300, 400, 1000\}; 
\item AOD: $[\max(-90^{o},n_{a}-\frac{w}{2}),\min(90^{o},n_{a}+\frac{w}{2})]$,
where $n_{a}\sim U([-90^{o},90^{o}])$, $w$ are uniformly random
selected from $\{10^{o},20^{o},30^{o},40^{o},100^{o}\}$. 
\item ZOD: $[\max(0,n_{a}-\frac{w}{2}),\min(180^{o},n_{a}+\frac{w}{2})]$,
where $n_{a}\sim U([0,180^{o}])$, $w$ are uniformly random selected
from $\{10^{o},20^{o},30^{o},40^{o},100^{o}\}$. 
\end{itemize}
We generate the test dataset $\mathcal{Y}$ using the UMa model with
20\% outdoor users and 80\% indoor users. Since the amplitude of CSI
data generated by the UMa model may be much smaller than that generated
by the CDL model, we normalize the test samples in $\mathcal{Y}$. 

\subsubsection{Datasets Used in Section\,\ref{sec:data_div}\label{APX: test data for diversity}}

The parameters listed in Appendix \ref{APX: test data for similarity}
are also used in the generation of the training and test datasets.
 For each training dataset, the path number, the time delay, AOD
and ZOD of each path in a sample are randomly generated from one of
the following ranges: 
\begin{itemize}
\item path number: $[1,n_{p}]$, where $n_{p}\in\{2,4,6,8,12,15,18\}$. 
\item time delay (ns): $[0,n_{t}]$, where $n_{t}\in\{200,800,1400,$ $2000\}$. 
\item AOD: $[-\frac{n_{a}}{2},\frac{n_{a}}{2}]$; ZOD: $[90^{o}-\frac{n_{a}}{2},90^{o}+\frac{n_{a}}{2}]$,
where $n_{a}\in\{80^{o},120^{o},$ $160^{o}\}$. 
\end{itemize}
\bibliographystyle{IEEEtran}
\bibliography{refs}

\end{document}